\documentclass[twocolumn,showpacs,preprintnumbers,amsmath,amssymb]{revtex4}

\usepackage{graphicx}
\usepackage{dcolumn}
\usepackage{bm}
\usepackage{booktabs}

\begin{document}
\preprint{APS}
\title{Two-photon excitation of rubidium atoms inside porous glass}
\author{L. Amy}
\author{L. Lenci}
\author{S. Villalba}
\author{H. Failache}
\email{heraclio@fing.edu.uy}
\author{A. Lezama} 
\affiliation{Instituto de F\'isica, Facultad de Ingenier\'ia, Universidad de la Rep\'ublica,\\ J. Herrera y Reissig 565, 11300 Montevideo, Uruguay}
\date{\today}

\begin{abstract}
We study the two-photon laser excitation to the 5D$_{5/2}$ energy level of $^{85}$Rb atoms contained in the interstices of a porous material made from sintered ground glass with typical pore dimensions in the  10 - 100 $\mu$m range. The excitation spectra show unusual flat-top lineshapes which are shown to be the consequence of wave-vector randomization of the laser light in the porous material. For large atomic densities, the spectra are affected by radiation trapping around the D2 transitions. The effect of the transient atomic response limited by time of flight between pores walls appears to have a minor influence in the excitation spectra. It is however revealed by the shortening of the temporal evolution of the emitted blue light following a sudden switch-off of the laser excitation.
\end{abstract}



\maketitle
\section{\label{Introduction}Introduction}

Light propagation in disordered systems is a subject of growing interest \cite{Wiersma:2013}. Motivated in particular by applications to medical imaging, a large attention is paid to the study of light propagation in highly scattering and turbid media \citep{Ishimaru:1978}. From a fundamental perspective, random media are studied in connection to Anderson's light localization \cite{Segev:2013} and as a possible support for random lasers \cite{Wiersma:2008}. Also, light propagation in random media has been suggested as a means to enhance spectroscopic sensitivity \cite{Svensson:2011,Redding:2013} and for energy storage \cite{Liu:2013} . 

The atoms from a dilute gas contained in the interstices of the random material can be seen as convenient local probes of the light inside the disordered medium. In addition, the atomic response to the light is affected, under suitable conditions, by the spatial confinement of the atomic motion imposed by the random material.  Such confinement effects can have spectroscopic consequences which in turn can be used to characterize the random medium \cite{Villalba:2013}. 

The spectroscopic study of atoms under confinement began with one dimensional (1D) confinement in vapor cells limited by flat transparent parallel windows with separations  of the order of a few to hundred microns \cite{Briaudeau:1996}. In these cells, due to the Doppler effect, laser excitation selects atoms with a well defined velocity component in the direction of confinement. Since the atom-light interaction is limited by the time of flight between opposing windows, the response is enhanced for atoms with a small velocity component in the direction of confinement resulting in sub-Doppler structures in the transmission spectrum \cite{Briaudeau:1996,Briaudeau:1999}.  More recently, the study of atoms in  thin cells was extended to sub-micron and even sub-wavelength cell-length and novel effects were reported such as the Dicke narrowing \cite{Sarkisyan:2004}.

Two dimensional (2D) atomic confinement was achieved and studied inside hollow optical fibers or photonics waveguides \cite{Benabid:2005,Yang:2007,Perrella:2013}. The limited time of flight of the atoms in the direction perpendicular to the fiber or waveguide axis can result in broadening of the excitation spectrum of the atoms by the fiber-guided light .

Few studies are concerned with atoms under 3D confinement. Ballin et al. \cite{Ballin:2013} recently reported the study of Cs atoms confined to the empty volume within a regular three-dimensional array of nanometric opals.  Our work, concerned with vapor contained in the micrometric interstices of a porous medium, can be seen as an implementation of a three-dimensional confinement.\\

The study of atoms inside disordered porous media was initiated by 
Svensson et al. \cite{Svensson:2011} who studied the absorption of O$_{2}$ molecules in a porous ceramic.  In previous work, we have addressed the spectroscopic study of rubidium vapor inside the  micrometric interstices of porous glass. We  studied first one-photon atomic resonances in the transmission spectrum  of the light exiting the porous sample after diffusion \cite{Villalba:2013} or backscattering from the incident surface  \cite{Villalba:2014}. Later, the nonlinear pump-probe spectroscopy was studied using two lasers in resonance with the Rb D1 lines \cite{Villalba:2014b}. In the present work we report the observation of two-photon excitation of $^{85}$Rb  atoms to the 5D$_{5/2}$ energy level inside a porous material.\\

Our sample is made by compaction of ground glass grains.  The shape of the interstices is random and highly irregular. Typical dimensions of the interstices are in the range 10 - 100  $\mu$m depending on the sample. Rubidium atoms are present as vapor inside the interstice volume and also in condensed phase on the glass surface of the pores. The gas phase density can be controlled through temperature.

The porous sample is an efficient light scatterer. Light propagation inside the porous medium is diffusive with a mean free path estimated to be a few tenth of a millimeter. After penetrating a distance of several mean free path, the light propagating inside the sample becomes a speckle field with random variations of amplitude, phase and polarization over distances of the order of the optical wavelength \citep{Goodman:2007}. Alternatively, the field inside the sample can be though as a superposition of plane waves with random wave-vectors, amplitudes and polarizations. 

Due to their thermal motion, during typical light-atom interaction time, the atoms travel distances that are larger than the optical wavelength thus exploring substantial changes in the optical field. Another consequence of the field randomization is that photons originated from the laser or emitted by the atoms cannot be distinguished by the direction of propagation since both experience similar diffusive propagation. Also, for sufficiently large vapor densities, the diffusive propagation of emitted photons through the sample favors re-absorption by resonant atoms leading to photon trapping \cite{Villalba:2013}.

A simplified energy-level scheme of $^{85}$Rb is presented in Fig. \ref{fsetup}. The  hyperfine level structure is not shown. For the excited states 5P$_{3/2}$ and 5D$_{5/2}$ the maximum hyperfine levels separations are 93 and 34 MHz respectively which is substantially  smaller than the Doppler width. Two photon excitation was achieved along the process: $5S_{1/2} \rightarrow  5P_{3/2} \rightarrow 5D_{5/2}$. The excited level $5D_{5/2}$  has a 239 ns  lifetime \cite{Sheng:2008}. Its main radiative decay path involves an infra-red  transition to levels $6P_{3/2}$  followed by the emission of blue light with 420 nm wavelength \cite{Heavens:1961}.\\

\section{\label{setup}Experimental setup}

\begin{figure}[h]
\includegraphics[width=9cm]{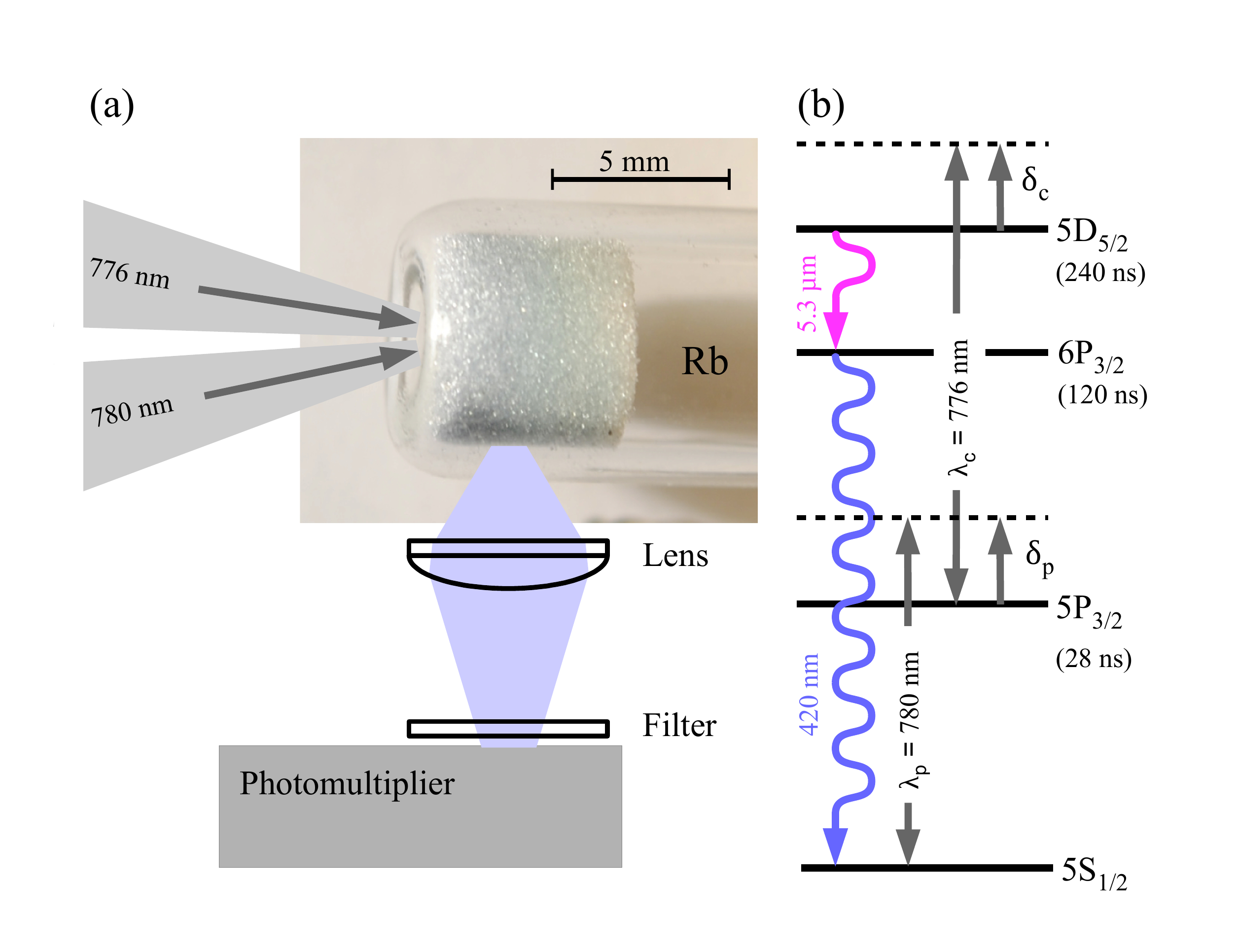}
\caption{(Color online) a) Experimental setup showing a porous cell. The lens images a portion of the porous medium surface on the photomultiplier window. The interferometric filter is centered around 420 nm with a 10 nm bandwidth. b) Atomic levels and optical fields involved in the experiment.} \label{fsetup}
\end{figure}

A picture of a porous sample is shown in  Fig.\ref{fsetup}(a). The samples were obtained from ground glass compacted and sintered at the flat end of a cylindrical 6 mm diameter glass tube. The size of glass particles was selected using sieves and decantatation columns. The cell was sealed under vacuum after introducing a small amount of condensed Rb through distillation. Three different cells were used in the experiments with typical grain sizes of 10, 50 and 100 $\mu$m. The actual statistical distribution of pore dimensions in our samples is yet unknown. We assume along this work that the typical pore dimensions are comparable to the corresponding typical grain size. Additional details on the cell fabrications are provided in  \cite{Villalba:2013,Villalba:2014}. During the experiments the cells were placed inside an oven for atomic density control by temperature variation.\\

We used two independent extended cavity diode lasers. A pumping laser, operating around the D2 lines of Rb (780 nm) and a coupling laser operating around the $5P_{3/2} \rightarrow 5D_{5/2}$ transition (776 nm). A frequency reference for the pumping laser was obtained from a saturation absorption setup. The coupling laser frequency was measured with a wavemeter with a resolution of $\pm$25 MHz.

The pump and coupling laser powers were 15 mW and 10 mW respectively. The laser beams were directed to the flat end of the sample glass tube.  Due to losses at the grain surfaces and leakage from the sample border, a strong decrease of the light intensity occurs as the light penetrates the porous medium.

Detection is achieved using a lens that produces the image of a portion of the sample lateral surface on a photomultiplier tube. The selected surface is separated from the flat end of the tube by several millimeters to ensure that the collected light had a diffusive propagation while traveling inside the sample. A blue transmitting dielectric bandpass filter (10 nm bandwidth) was placed at the entrance of the photomultiplier tube to select the 420 nm emission and reject laser light.\\ 

The excitation spectra were recorded by tuning one laser while keeping the second laser frequency constant. In addition, we have studied the transient decay of the emission at 420 nm arising when both lasers are tuned to resonance. For this the light from one laser was suddenly turned  off with the help of an optical fiber Mach-Zender modulator (switching time $\lesssim 10$ ns).

\begin{figure*}
\includegraphics[width=1\textwidth]{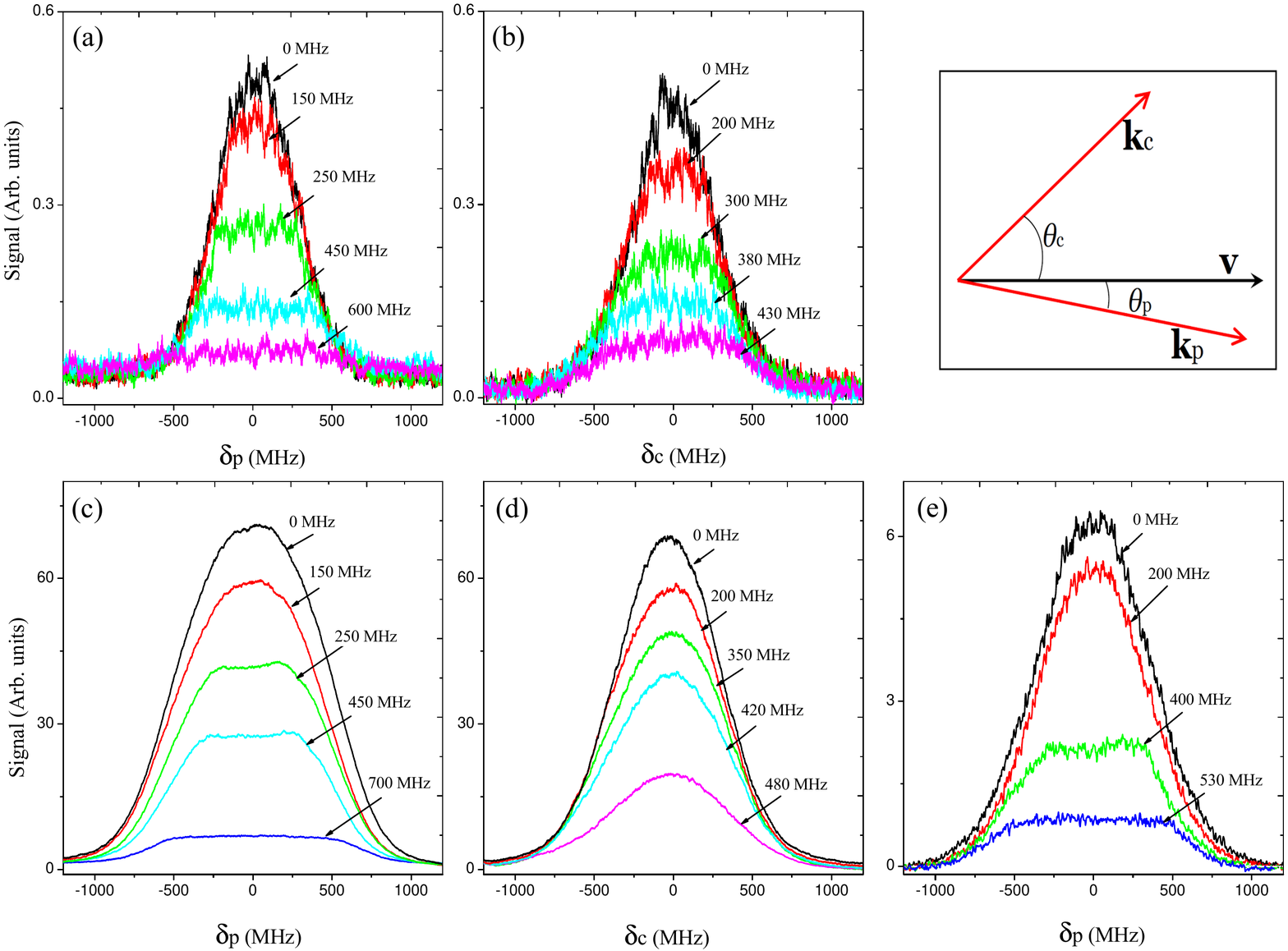}
\caption{(Color online) Two-photon excitation spectra obtained by tuning the frequency of one laser while keeping the second laser frequency fixed. The detuning of the fixed frequency laser is indicated for each plot. Upper row: T $\approx$ 40$^{\circ}$C. Lower row: T $\approx$ 120$^{\circ}$C. (a-d): cell with 100 $\mu$m typical pore size, (e): cell with 10 $\mu$m typical pore size. Inset: atomic velocity and two randomly oriented pump and coupling photons wavevectors.} \label{exp1}
\end{figure*}

\section{\label{model}Excitation spectra}

Figure \ref{exp1} shows the two-photon excitation spectra obtained scanning either the pump or the coupling laser for different detunings of the fixed frequency laser. In these plots the zero of the detuning axis correspond to the position where the maximum signal is observed after freely varying both lasers frequencies. These line-centers are located approximately at the maximum of the Doppler broadened absorption line of the corresponding transition (including the unresolved hyperfine structure). The upper row in Fig. \ref{exp1} correspond to spectra taken at low atomic density (T $\approx$ 40 $^{\circ}$C) while the lower row correspond to large atomic density (T $\approx$ 120 $^{\circ}$C).\\ 

A first noticeable property of the spectra in Fig. \ref{exp1} is their symmetry with respect to zero detuning (a minor asymmetry can be attributed to the unresolved   hyperfine structures). This fact is strikingly different from what occurs in ordinary vapor cells where, in general (depending on the angle between laser beams), the spectra are asymmetric with respect to the atomic resonance frequency and presents maxima whose positions vary according to the detuning of the coupling field (see for instance \cite{Bjorkholm:1976}).\\
   
A second remarkable feature in the spectra, visible in Figs. \ref{exp1}(a-c) and \ref{exp1}(e), is the occurrence of flat-top lineshapes whose width is approximately given by twice the absolute value of the detuning of the coupling laser.\\

We interpret these two properties as the consequence of the field randomization inside the porous material. A simplified description of two-photon excitation identifies two different excitation channels \cite{Berman:2010}: A \emph{stepwise} excitation in which the atom is first promoted to the intermediate state from which it is excited to the final state. A \emph{direct} excitation in which the atom is directly promoted from the ground level to the final level via the absorption of a photon from each field whose total energy equals the energy separation between initial and final levels. While the coherence between fields is not required for the \emph{stepwise} process, it is an essential condition for the \emph{direct} mechanism. Field randomization is responsible for the rapid variation of the fields relative phase during the light atom interaction process. This results in strong inhibition of the \emph{direct} excitation channel.\\

Neglecting the homogeneous linewidth of the atomic transitions, for an atom moving with velocity $\mathbf{v}$ in the presence of two fields with wavevectors $\mathbf{k_p}$ and $\mathbf{k_c}$ the resonance conditions for the \emph{stepwise} process are:
\begin{subequations}\label{step}
\begin{eqnarray} 
 \delta_{p}\simeq\mathbf{k_p}\cdot \mathbf{v}=k_p v \cos (\theta_p) \label{pump}\\
  \delta_{c}\simeq\mathbf{k_c}\cdot \mathbf{v}=k_c v \cos (\theta_c) \label{coup}
 \end{eqnarray} 
 \end{subequations}
where the detunings $\delta_{p}$ and $\delta_{c}$ are defined in Fig. \ref{fsetup} and $\theta_p$ and $\theta_c$  refer to the angles between the atom velocity and the corresponding wavevector (see inset in Fig. \ref{exp1}).\\

For the \emph{direct} process the resonance condition is:  
\begin{eqnarray}
 \delta_{p}+ \delta_{c}\simeq(\mathbf{k_p}+\mathbf{k_c})\cdot \mathbf{v} \label{direct} 
 \end{eqnarray}

If the detuning $\delta_{c}$ is fixed, in the \emph{stepwise} process the atomic velocity is selected though condition \ref{coup}. For a selected velocity,  the values  of $\delta_{p}$ that satisfy condition \ref{pump} have an even distribution around $\delta_{p}=0$ as a consequence of the isotropic random orientation of the wavevector $\mathbf{k_p}$ in the speckle field. 

On the other hand, for the  \emph{direct} process (Eq. \ref{direct}) the wavevector randomization leads to a distribution of $\delta_{p}$ centered around $-\delta_{c}$. Analogous arguments apply if the roles of the pump and coupling fields are reversed.

The fact that the observed spectra are centered on the atomic resonance transition and that there is no observable peak opposite to the detuning of the fixed-frequency laser is a strong indication that the \emph{direct} process is inhibited inside the porous medium.\\

A simple qualitative explanation can be given as follows for the plateau-like spectra observed in Figs. \ref{exp1}(a-c,e) \cite{Carvalho:2015}. Considering $\delta_{c}$  as a fixed detuning, $\vert\delta_{c}\vert/k_c$ determines the minimum velocity for the verification of Eq. \ref{coup}. For velocities above this minimum, for any value of $\vert\delta_{p}\vert$  smaller than $k_p\vert\delta_{c}\vert/k_c$, Eq. \ref{pump} can be fulfilled for some angle $\theta_p$. The plateau originates from the uniform distribution probability for $\cos(\theta_p)$. The plateau ends when  $\vert\delta_{p}\vert = k_p\vert\delta_{c}\vert/k_c$. For larger values of $\vert\delta_{p}\vert$,  Eq. \ref{pump} determines the minimum required atomic velocity thus reducing the number of atoms participating in the process as $\vert\delta_{p}\vert$ is increased.\\

For the two-photon \emph{stepwise} excitation process, the excitation probability $S_0(\nu_p,\nu_c)$ for an atom traveling with velocity $\mathbf{v}$ in the presence of laser fields with wavevectors $\mathbf{k_p}$ and $\mathbf{k_c}$ is given by:
\begin{eqnarray}
S_0(\nu_p,\nu_c)\propto F_p( \nu_p-\dfrac{v}{\lambda_p}\cos\theta_p) F_c( \nu_c-\dfrac{v}{\lambda_c}\cos\theta_c) \label{Scero}
\end{eqnarray}
where $\nu_{p(c)}$, $\lambda_{p(c)}$ and $F_{p(c)}(\nu)$ represent respectively the frequency, wavelength and  normalized absorption profile in the atomic reference frame for the field and transition corresponding to the pump(coupling) field. 

In vapor cell spectroscopy the functions $F_{p(c)}(\nu)$ are usually considered to be homogeneous over the atomic ensemble. However, one has to keep in mind that for atoms confined in small volumes these functions could depend on the atomic velocity and position. Initially neglecting such dependencies, the excitation spectrum profile can be obtained by summation over all possible atomic velocities and all orientations of vectors $\mathbf{k_p}$ and $\mathbf{k_c}$.
\begin{align}
& S(\nu_p,\nu_c)\propto \nonumber \\ 
&\int_0^{\pi}\sin\theta_p d\theta_p\int_0^{\pi}\sin\theta_c d\theta_c\int_0^{\infty}\mathcal{V}(v)S_0(\nu_p,\nu_c)dv \label{integral}
\end{align}
where $\mathcal{V}(v)$ is the probability distribution of the atomic velocity modulus.\\

An equation similar to Eq. \ref{integral} has been previously encountered in the context of radiation trapping and near resonant frequency redistribution in an atomic vapor \cite{Molisch:1998}. In that context, $\nu_{p}$ and $\nu_{c}$ refer to the frequencies of the absorbed and re-emitted photons respectively. Similarly to our case where photons with random polarization arrive from random directions, radiated photon are emitted in a random direction (provided that polarization is not considered). The two problems are thus intimately connected. However, in the case of radiation redistribution, Eq. \ref{Scero} is replaced by: 
\begin{align}
S_0(\nu_p,\nu_c)&\propto F_p( \nu_p-\dfrac{v}{\lambda_p}\cos\theta_p) \nonumber \\ 
&\times p( \nu_c-\dfrac{v}{\lambda_c}\cos\theta_c,\nu_p-\dfrac{v}{\lambda_p}\cos\theta_p) \label{Scero_2}
\end{align}
where $p(\nu,\nu^{\prime})$ is the probability for emission of a photon of frequency $\nu$ after absorption of a photon of frequency $\nu^{\prime}$ in the atomic rest frame.

Equation \ref{Scero_2} reduces to the form of Eq. \ref{Scero} in two particular cases: a) Infinitely narrow transition and perfectly elastic light scattering [$F_p(\nu)=\delta(\nu-\nu_0), p(\nu,\nu^{\prime})=\delta(\nu-\nu^{\prime})$]. b) Uncorrelated frequencies in the atomic  reference frame  for emitted and absorbed photons. These two cases are discussed in  \cite{Molisch:1998} and for case a) an analytical expression for $S(\nu_p,\nu_c)$ is given.\\
  
\begin{figure}[h]
\includegraphics[width=1\linewidth]{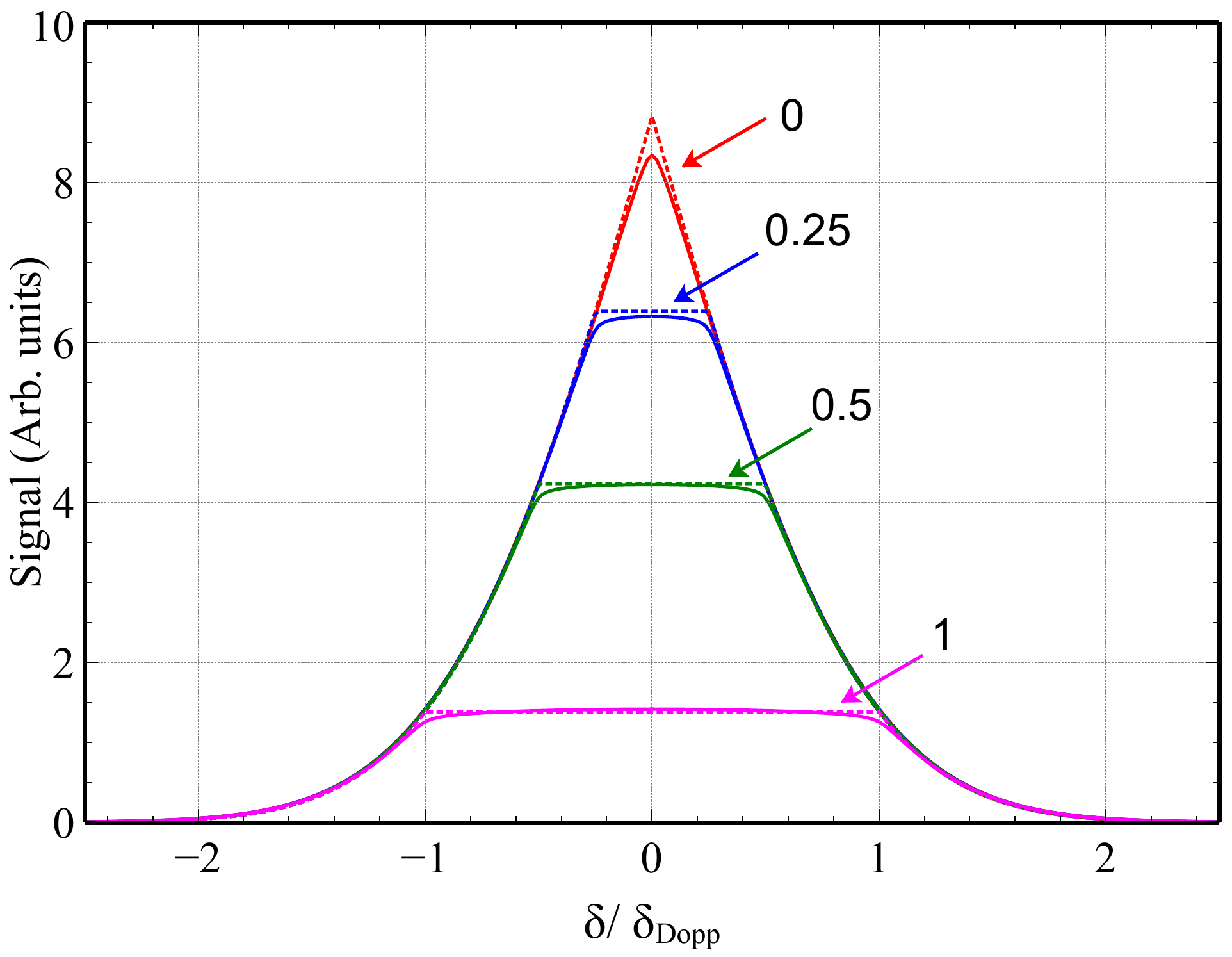}
\caption{(Color online) Calculated excitation spectra as a function of the scanned laser detuning calculated using  Eq. \ref{integral} for different values of the fixed-frequency laser detuning.. Both detunings are expressed in units of the corresponding Doppler width $\delta_{Dopp}\equiv \sqrt{2k_BT/m}/\lambda$. Dashed: negligible absorption width.  Solid: Lorentzian absorption lineshapes with linewidth determined by the 5P$_{3/2}$ radiative decay rate (6 MHz).} \label{thed}
\end{figure}

The spectrum profiles calculated using Eq. \ref{integral} are shown in Fig. \ref{thed}. In this figure the laser detunings are expressed in units of the corresponding Doppler detuning $\delta_{Dopp}\equiv \bar{v}/\lambda$ where $\bar{v}\equiv\sqrt{2k_BT/m}$ is the most probable atomic velocity modulus ($\bar{v}\approx$  240 m/s at room temperature corresponding to $\delta_{Dopp}\approx$  310 MHz for $\lambda=$ 780 nm). The dotted lines correspond to the assumption of infinitely narrow absorption profiles for the lower and upper atomic transitions. Notice the perfectly flat top of the spectra and the singularities in the spectral function derivative occurring for $\vert\delta_{p}\vert \lambda_p= \vert\delta_{c}\vert \lambda_c$. A better approximation to the experimentally observed lineshapes is obtained when the width of the atomic absorption line is considered. The plots shown in Fig. \ref{thed} (solid) assume Lorentzian absorption lineshapes with width determined by the intermediate level (5P$_{3/2}$) radiative decay rate (6 MHz). The flat tops of the spectral lines are now slightly rounded and the singularity in the derivative is eliminated. Keeping in mind  that the unresolved hyperfine structure introduces additional broadening in the experimental spectra, a satisfactory agreement is observed with the theoretical simulations. 

As already mentioned, the assumption of homogeneous atomic absorption spectral profiles can be questioned for atoms under confinement. We have numerically evaluated the excitation spectra considering the transient atomic response of the atoms starting when the atom leaves a pore wall and ending when it collides with another  wall. Such transient evolution introduces a velocity and position dependency of the atomic response. Our computation was based on assumptions about the statistical distribution of atomic trajectories inside the pores. However, for all reasonable assumptions consistent with the estimated pore sizes, the effect of the transient evolutions on the excitation lineshape appears to be minor and is not presented here.\\

It is worth mentioning that in the context of the theory of radiation trapping the curves plotted in Fig. \ref{thed} represent an idealized process corresponding to a \emph{single} absorption-radiation cycle \cite{Carvalho:2015}. Experimental verification of these predictions are elusive or indirect since in actual radiation trapping experiments \emph{multiple} absorption-radiation cycles do occur and significantly modify the spectrum. Our result can be seen as a first experimental test of the predictions corresponding to the single absorption-radiation cycle.\\  

Figures \ref{exp1}(a) and \ref{exp1}(b) correspond to relatively low atomic density (T $\sim$ 40 $^{\circ}$C). Under these conditions the excitation spectra obtained by exchanging the fixed and scanning frequency lasers show similar features including the flat-top spectra. Notice however that the spectra obtained when the coupling laser is scanned [Fig.\ref{exp1}(b)] are not as neatly flat. This asymmetry under exchange of the roles of the two laser fields is not compatible with Eq. \ref{integral}. The asymmetry becomes  more evident as the density of the sample is increased. Figures \ref{exp1}(c) and \ref{exp1}(d) were taken at a temperature T $\sim$ 120 $^{\circ}$C.  The two sets of spectra are now clearly different. Flat-top spectra are obtained when the pump laser is scanned and an overall broadening is observed. On the other hand bell shaped lines with no noticeable dependency on the pump frequency position are observed when the coupling laser is scanned. We interpret this asymmetric behavior as the consequence of the onset of photon trapping.

Photon trapping around the resonant $5S_{1/2}(F=2) \rightarrow 5P_{3/2}$ transition is known to occur in our samples when the atomic density is sufficiently large \cite{Villalba:2014}. On the other hand, negligibly photon trapping is expected to occur around the $5P_{3/2} \rightarrow 5D_{5/2}$ transition due to the small population in the $5P_{3/2}$ level. 

When photon trapping occurs around the lower transition, an additional path for two-photon excitation is possible involving one photon re-emitted around the $5S_{1/2} \rightarrow 5P_{3/2}$ transition and one photon from the coupling laser. If the pump laser frequency is not too far detuned from  the Doppler broadened atomic absorption line and for large enough atomic densities the spectrum of the trapped photons closely approaches the absorption profile with little dependency on the pump laser detuning \cite{Molisch:1998}. In consequence, when the pump laser frequency is scanned the contribution to the spectrum of the two photon-excitation mechanism involving re-emitted photons remains approximately constant until the pump field detuning is sufficiently large ($\vert\delta_{p}\vert \gtrsim 2 \delta_{Dopp}$ \cite{Molisch:1998}). We attribute the spectral broadening observed in all spectra in  Fig. \ref{exp1}(c) to the contribution of trapped photons to the excitation of level 5D$_{5/2}$. Conversely, when the coupling laser is scanned, the excitation mechanism involving trapped photons results in a contribution to the spectrum that roughly follows the spectral density of the trapped photons which is essentially independent from the pump field detuning [Fig. \ref{exp1}(d)].\\

The features appearing in the excitation spectra described above show variations over a typical scale of the order of $\sim$100 MHz. On the other hand the frequency scale associated to the atomic flight between pore walls is of the order of $\bar{v}/L$ where $L$ is the characteristic pore size. For our samples with typical pore sizes in the 10 - 100 $\mu$m range, the frequency scale associated to confinement corresponds to 2 - 20 MHz. In consequence, the effect of the transient atomic evolution due to confinement is expected to have small influence on the excitation spectra. This expectation is experimentally supported by the similarities between  the spectra taken with the 100 $\mu$m pores cell (Fig. \ref{exp1}(c)) and with the 10 $\mu$m pores cell (Fig. \ref{exp1}(e)) when the pump laser is scanned. An overall narrowing of the spectra in Fig. \ref{exp1}(e) compared to Fig. \ref{exp1}(c) can be attributed to a reduced contribution of the two-photon excitation mechanism involving trapped photons \cite{Villalba:2013}.\\

\section{\label{time}Transient signal decay}

In order to detect the effect of spatial confinement on the atomic response, we have studied the temporal decay of the emitted blue radiation after a sudden switch-off of the coupling laser using a porous cell with typical pore size 50 $\mu$m. In these measurements the temperature of the cell was kept at T = 120$^{\circ}$C and both lasers were tuned to line-center ($\delta_p=\delta_c=0$).

In the absence of any spatial confinement effect, sufficiently long after the turning off of the coupling laser, the blue fluorescence is expected to decay exponentially with a decay time determined by the $5D_{5/2}$ level lifetime (239 ns) \cite{Sheng:2008}.

\begin{figure}[h]
\includegraphics[width=1\linewidth]{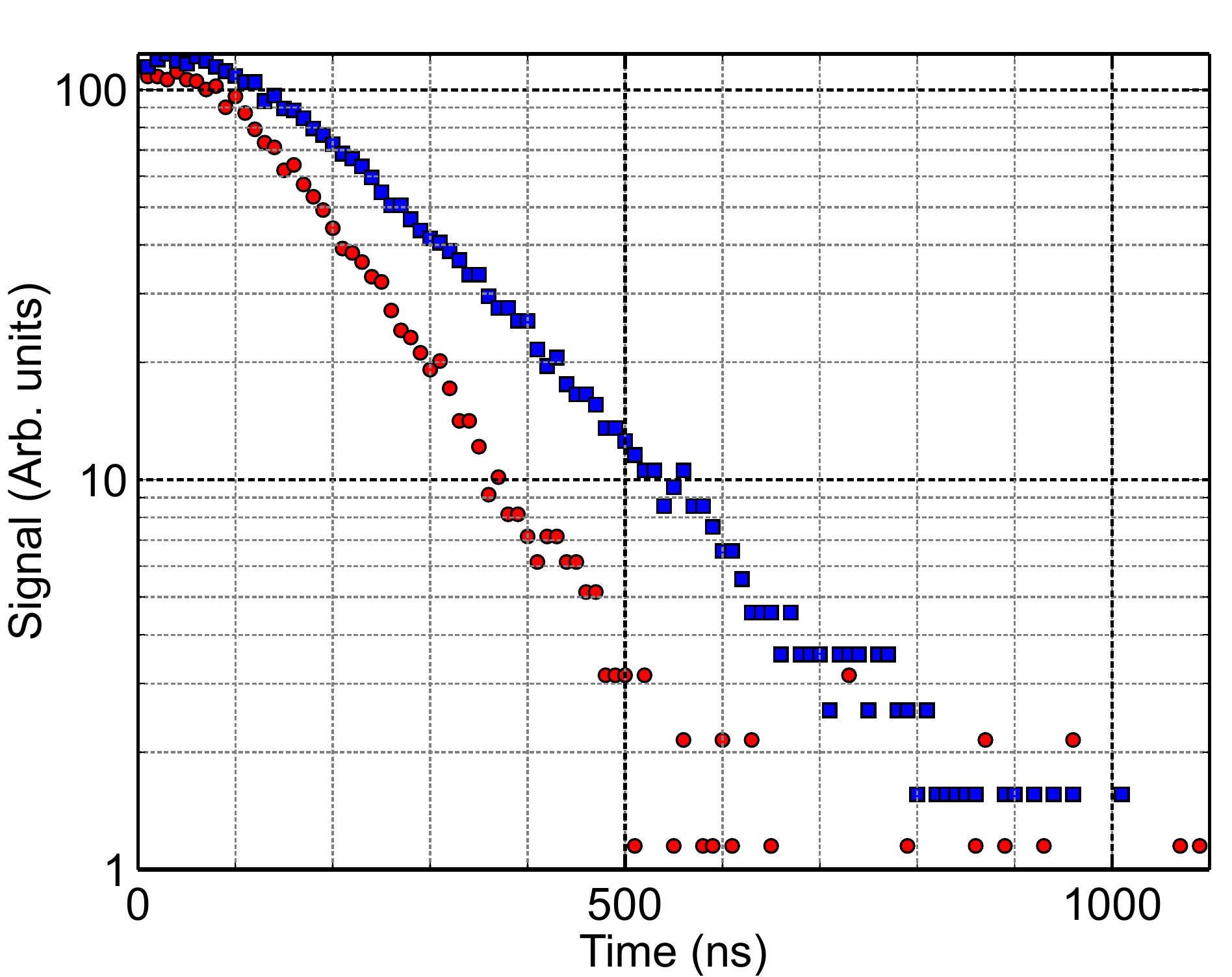}
\caption{(Color online) Temporal evolution of the blue fluorescence after turn-off of the coupling laser. Blues squares: light from unconfined vapor. Red circles: light form porous sample.} \label{exp3}
\end{figure}

We initially recorded the transient signal originating from the portion of the glass tube that is not occupied by the porous material (See Fig. \ref{fsetup}). For long times we measure an exponential decay time $\tau_u =$ 180$\pm$10 ns (Fig. \ref{exp3}). We attribute this reduced decay time to collisions with contaminants present in the Rb vapor.

We have next measured the long-time decay time for the blue light originating from the porous medium. A shorter decay time $\tau_c =$ 120$\pm$10 ns is observed. Since the gas composition is expected to be the same inside and outside the porous medium, we attribute this additional shortening to the interruption of the emission process by the collisions of the atoms to the cell walls. From the two observed decay times we deduce a characteristic time $\tau_F \approx$ 360$\pm$100 ns for the atomic time of flight between collisions with the pore walls [$\tau_F\equiv(\tau_c^{-1}-\tau_u^{-1})^{-1}$].  Considering the typical dimension of the pores ($\sim$50 $\mu$m) the typical time of flight corresponds to a velocity $v_F\approx 140\pm40$ m/s. This velocity is smaller than the mean velocity $\bar{v}$  which suggests that the two-photon excitation involving diffuse fields could favor slow atoms.    

\section{Conclusions}

We have observed and studied the two-photon excitation of an alkali vapor confined to the micrometric-size interstices of a porous glass. Due to the diffuse nature of the light fields, the two-photon excitation is essentially due to a \emph{stepwise} process involving the excitation of the intermediate state. In consequence, the excitation spectra are centered around the atomic transitions. In addition the field wavectors randomization results in singular plateau-like line-shapes similar to those predicted for the spectrum of the light emitted from a resonant atomic vapor in a single absorption-emission cycle. A significant reduction on the fluorescence decay time was measured as a consequence of the spatial confinement of atoms excited to a long-lived excited state. Further investigation is required to explore velocity selection in two-photon excitation by diffuse fields.\\

The cells considered in this work have typical pore dimensions which are larger than optical wavelength. At present,  new porous samples are being prepared with sub-micrometer pore sizes. The radiation spectra of atoms contained in the interstices of these samples should reveal new features corresponding to a regime in which the effects of confinement are dominant.  

\section{Acknowledgments}

We wish to thank P. Valente, D. Bloch, M. Chevrollier, M. Ori\'{a} and A. Laliotis for fruitful discussions. We acknowledge financial support from ECOS-Sud, CSIC and CAP (Universidad de la Rep\'ublica).\\


\end{document}